\documentclass[12pt,twoside,a4paper,useAMS]{meteoroids2013}
\usepackage{color,graphicx}
\usepackage[english]{babel}

\title[Pro tempore showers in EDMOND] 
{Confirmation and characterization\\ of IAU temporary meteor showers\\ in EDMOND database} 

\author[Korno\v{s}, Matlovi\v{c}, Rudawska, T\'{o}th, Hajdukov\'{a}, Koukal, Piffl]   
{Korno\v{s} L.$^1$, %
 Matlovi\v{c} P.$^1$, %
 Rudawska R.$^1$, %
 T\'{o}th J.$^1$, %
 Hajdukov\'{a} M.~Jr.$^2$, %
 Koukal J.$^3$,  
 \and Piffl R.$^3$}

\affiliation{$^1$Faculty of Mathematics, Physics and Informatics, Comenius
University in Bratislava, \break Slovak Republic (email: kornos@fmph.uniba.sk)
\\[\affilskip]
$^2$Astronomical Institute of the Slovak Academy of Sciences, Bratislava, Slovak
Republic
\\[\affilskip]
$^3$CEMeNt - Central European Meteor Network }

\pagerange{110--116}
\jname{Proceedings of the Meteoroids 2013 Conference\\
       Aug. 26-30, 2013, A.M. University, Pozna\'{n}, Poland}
\editors{Jopek T.J., Rietmeijer F.J.M., Watanabe J., Williams I.P., ed.}
\begin{document}
\maketitle
\begin{abstract}
The European viDeo MeteOr Network Database (EDMOND) is a database of video meteor
orbits resulting from cooperation and data sharing among several European national
networks and the International Meteor Organization Video Meteor Network, IMO VMN,
(\cite{Kornos2013b}). At present, the 4th version of the EDMOND database, which
contains 83\,369 video meteor orbits, has been released.

The first results of the database analysis, in which we studied minor streams, are
presented. Using the radiant-geocentric velocity method we identified 267 meteor
showers, among them 67 established showers and 200 from the working list of the IAU
MDC. Making a more detailed examination, we clearly identified 22 showers of 65
\textit{pro tempore} showers of the working list of the IAU MDC (updated in August
2013). The identification of 18 meteor showers was questionable, while 25 showers
were not found. For all the identified temporary meteor showers, we list the
weighted mean orbital elements, the radiant position and the geocentric velocity.
\end{abstract}
\section{Introduction}
The rapid development of video techniques in recent years has resulted in the
massive use of video cameras in meteor observations. The number of new meteor
networks has increased, and the efficiency of those already existing has improved.
In three years, the Japanese meteor network database, containing around 30 low-light
level camera observations, grew to 65\,000 orbits (\cite{SonotaCo2009};
\cite{SonotaCo2010}). The recently  established system CAMS (Cameras for Allsky
Meteor Surveillance) in the United States obtained 47\,000 orbits of meteors just in
the first year of its operation (\cite{Jenniskens2011}). In Europe,  between 2000
and 2013, the IMO Video Meteor Network collected over 1.2 million single-station
meteors (\cite{Molau2014}). Also, in Europe, the continuous monitoring of meteors
and fireballs is conducted by the 25 stations the Spanish Meteor and Fireball
Network (SPMN; \cite{Pujols2013}), which has been working now for 5 years. While
till mid-2011, NASA's All-Sky Fireball network, established in 2008, with its 6
video cameras, detected 1796 multi-station meteors (\cite{Cooke2012}). Another good
example of well developed regional networks are the Canadian Automated Meteor
Observatory (CAMO; \cite{Brown2010}) and the Croatian Meteor Network (CMN;
\cite{Andreic2010}).

Thanks to the broad international cooperation of video meteor observers from several
European countries, a multi-national network EDMONd (European viDeoMeteor
Observation Network) was created. As a result of its work, the first version of the
EDMOND database, containing data from the years 2009, 2010, 2011 and half of 2012,
was presented at the IMC conference in La Palma, Spain in 2012 (\cite{Kornos2013a}).
In the last year, observers affiliated to the International Meteor Organization
Video Meteor Network (IMO VMN) have started to share their data, whereas the data of
EDMONd and IMO VMN have been merged. Nowadays, the data is collected from observers
from a substantial part of Europe and, due to this international cooperation, meteor
activity is monitored over almost the entire Europe. In effect, the database has
accumulated 1\,639\,358 records of single-station meteors between 2000 and 2013
(EDMONd -- 447\,266 and IMO VMN -- 1\,192\,092).
\section{EDMOND database}
The computation of meteor orbits is performed by the UFOOrbit software
(\cite{SonotaCo2009}). As the single-station video data are obtained and reduced
using two different tools, the MetRec (\cite{Molau1999}) and UFOAnalyzer tools
(\cite{SonotaCo2009}), the UFO data can be used without any changes. However, the
data obtained by the MetRec software has to be first converted into the UFO format
using the program INF2MCSV written by SonotaCo. The present database contains about
72 \% of MetRec data. As the conversion is not fully compatible, the computation of
orbits is performed in two steps. First, preliminary orbits are computed using
UFOOrbit with basic parameter settings $Q_o$ and $dt=5$ sec (which means that all
combinations of single-station meteors within 5 second intervals are computed), and
with additional settings: beginning and terminal heights have to be
$H_{1,2}\in(15;200)$\,km, the empirically derived quality parameter $QA>0.3$, and
the largest difference in velocity among considered stations in the orbit
computation is $dV<7$\,km/s.

After that, to reject the less precise orbits and false orbits, another filter of
parameters is applied: the angle of observed trajectory has to be $Q_{o}>1$\,deg,
the duration of the meteor $dur>0.1$\,sec, the convergence angle $Q_{c}>10$\,deg,
the difference between the two poles of ground trajectory $dGP<0.5$\,deg, and the
difference in velocity between unified velocity and velocity from one of the
stations $dv12\%<7.07\,\%$. In comparison to the previous versions of the database,
the most important modification is the restriction of the difference in velocities
for stations used in meteor orbit computation. The definition of all parameters is
in the UFO Manual. More details can be found in (\cite{Kornos2013b}).

At present, the 4th version of the EDMOND database containing 83\,369 video meteor
orbits, has been released. Most of them ($\sim$\,84\%) are double-stations orbits.
About 48\,800 orbits belong to the sporadic background and 34\,500 are shower
meteors (59\,\% and 41\,\%, respectively).

The EDMOND database was examined in several tests and compared with other meteor
orbits databases. The examination allowed us to demonstrate the characteristic
features of the EDMOND, which are particularly important for future analyses based
on the data.

The derivation of orbital elements, which define the shape of the orbit, is highly
dependent on the uncertainty of the determination of the meteor velocity. One of the
parameters used in the data reduction is dv12\% (the difference in the velocities,
given in percentage). The geocentric velocity is the decisive parameter in the
calculation of the orbit. Thus, the dv12\% parameter is an important indicator of
its accuracy. The smaller the difference between velocities from different stations,
the more accurate the orbit determination is. Therefore, in the distribution of
dv12\%, a decrease in the number of orbits with increasing values of dv12\% should
be the most rapid. The comparison of the distribution of dv12\% parameter of the
EDMOND (\cite{Kornos2013a}) and SonotaCo catalogue showed similar decrease; with a
slightly slower one in the EDMOND.

\begin{figure}
\centerline{\includegraphics[width=6.0cm]{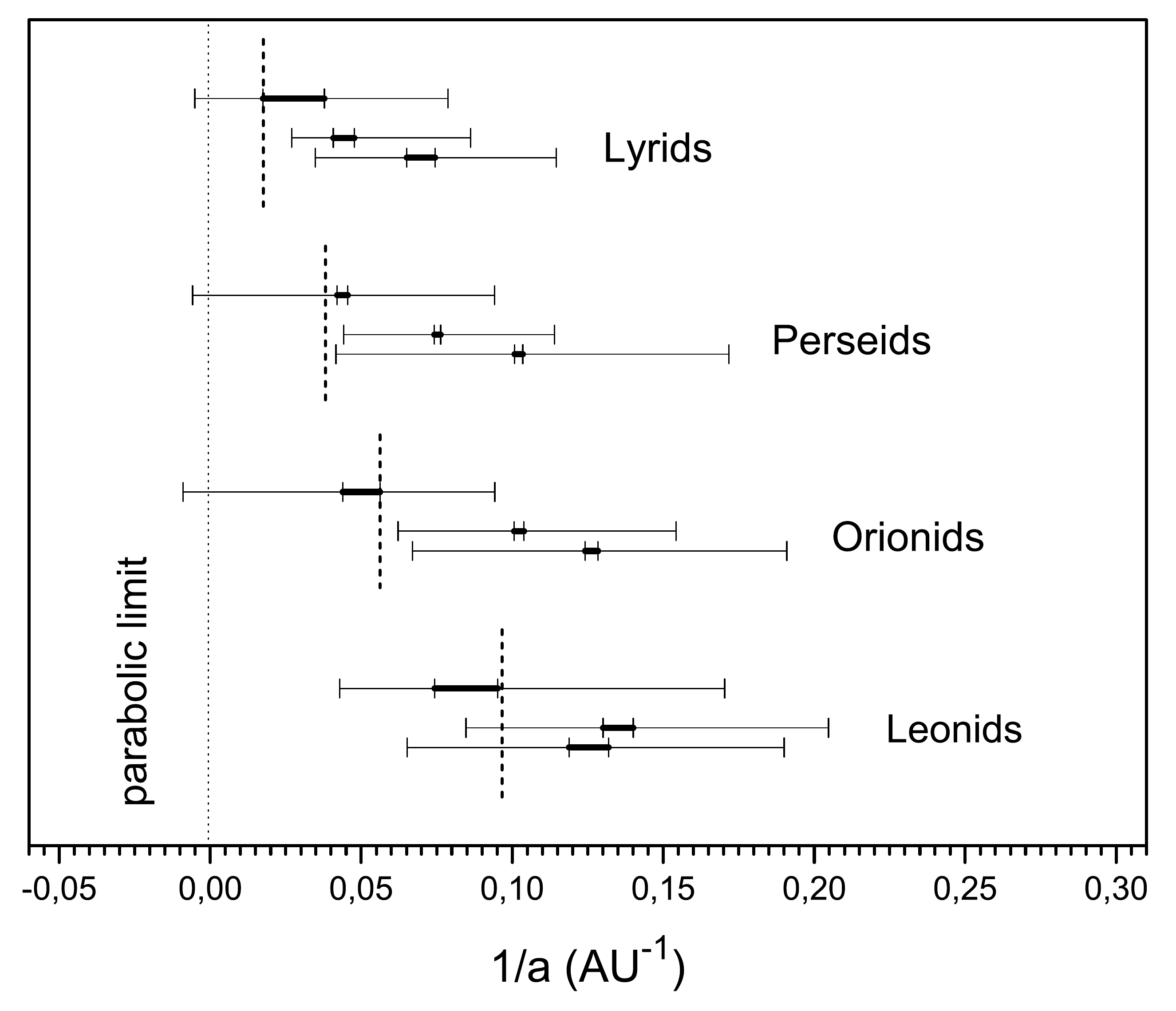}}
\footnotesize
\caption{ Comparison of the observed dispersion for the chosen meteoroid streams
from different databases (upper line - photographic meteors from the IAU MDC, middle
line - video meteors from the SonotaCo catalogue (2007 - 2009), lower line - the
EDMOND data) described by absolute median deviation in terms of $1/a$: Thin line -
the interval between two limiting values of $(1/a)_{1/2}$, which includes 50 percent
of all orbits. Bold line - the interval between two limiting values of the
uncertainty $(1/a)_L$ of the resulting values of median $(1/a)_M$. Dotted vertical
line - the parabolic limit. Dashed vertical lines - parent comets (the figure is
taken from the paper (\cite{Hajdukova2013}).} \normalsize \label{fig1}
\end{figure}

The distributions of orbital parameters within several meteoroid streams from EDMOND
were also analysed. In~\cite{Kornos2013a}, the dispersions of orbital elements of
the Lyrids from EDMOND were studied. Comparing them with the SonotaCo video orbits,
the consistency of both datasets was demonstrated. Moreover, if we compare both sets
of video data (Figure~\ref{fig1}), the dispersions in $1/a$ of the meteor orbits
within individual streams obtained from the EDMOND data are about 1.3 times larger
than the values from the SonotaCo catalogue. Figure~\ref{fig1} shows the observed
differences in the semi-major axes within the meteor streams compared with the
orbital deviations in the streams determined from different datasets. The median
semimajor axes of video meteor orbits in both the EDMOND and SonotaCo data are
systematically biased, probably as a consequence of the method used to determine the
orbits. In comparison with the IAU MDC photographic database (\cite{Lindblad2005}),
they are shifted towards the short-period side; the velocities determined in the
video data are slightly underestimated.

An important indicator of the quality of data is the relative number of hyperbolic
orbits, because the probability of registering real hyperbolic orbits is very small
(\cite{Hajdukovaetal2014a}). The apparent hyperbolicity of the orbits is, generally,
caused by a high spread in velocity determination, shifting a part of the data
through the parabolic limit. This, however, does not explicitly mean large
measurement errors. Of the 83\,369 meteor orbits collected in the EDMOND, 5.7\% are
determined as hyperbolic. This percentage is roughly comparable to that in the
SonotaCo database. Initially, the proportion of hyperbolic meteors in the latter was
11.58\%, but after the selection of quality orbits (\cite{Veres2010}), this was
reduced to 3.28\%. Of the 4712 hyperbolic meteors in the EDMOND, 43\% are shower
meteors. Shower meteors which have heliocentric velocities with excesses over the
parabolic limit offer proof of the false hyperbolicity of their orbits. The
hyperbolic orbits in our data were analysed separately in the
paper~\cite{Hajdukova2014b}.

A comparison of both the EDMOND and the SonotaCo catalogue, in terms of orbital
parameters, showed an equivalence of the data.

\section{Identification of streams in EDMOND}
Meteor showers in the EDMOND database were identified using the IAU Meteor Data
Center Database (IAU MDC; \cite{Jopek2014}). At the end of August 2013, the IAU MDC
list of showers contained 461 showers, 95 of them established and 366 in the working
list.

In the first search, the radiant position-geocentric velocity method was used (we
hereafter call it \textit{radiant-$V_g$} method). Meteors were selected according to
the peaks activity of meteor showers ($\pm$15~deg) given in the IAU MDC list, and
fulfilling the conditions for radiant position ($\pm$5~deg) and geocentric velocity
($\pm$10\%$\cdot V_g$). A shower was considered only if at least 5 orbits had been
identified. In this way, 267 meteor showers were identified, where 200 of them are
meteor showers from the working list and 67 are the established showers.

We focused on \textit{pro tempore} showers from the IAU MDC working list. Of 65
\textit{pro tempore} showers in the list, 61 were identified using the
\textit{radiant-$V_g$} method. To determine their fundamental parameters more
precisely, the first part of the Welch method (\cite{Welch2001}) with
Southworth-Hawkins $D$ criterion (~\cite{SouthworthHawkins1963}) was used. According
to the equation (4) in \cite{Welch2001} paper

\begin{equation}
\rho = \sum_{i=1}^{N} \left(1-\frac{D_i^2}{D_c^2}\right)~; %
  \quad\quad
  D_i\leq D_c~,
\label{eqn1}
\end{equation}

\noindent where $\rho$ is a density at a point in orbital elements space, $N$ is the
number of meteors of a \textit{pro tempore} shower found in the first step, $D_i$ is
the value obtained for the $i$-th meteor in the \textit{pro tempore} shower by
comparing its orbit with orbits of each member of the identified shower, and $D_c$
is the threshold value that determines the dynamical similarity among meteor orbits.
We searched for the core of each \textit{pro tempore} shower identified in the first
step of the analysis (i.e. by \textit{radiant-$V_g$} method).

The procedure creates a group of meteors around each meteor orbit from the examined
shower, which fulfil the condition of the limiting value of Southworth and Hawkins
criterion $D_c=0.12$. On the basis of the equation (\ref{eqn1}), the value of the
density ($\rho$) is determined for each group. The higher the density value, the
more important the group in the examined shower is. However, the highest value of
$\rho$ does not always mean it is the core of the stream because the initial set
could be contaminated by a nearby separate small shower; or because the MDC data are
not yet accurate enough.

We therefore compared all the available parameters of each \textit{pro tempore}
shower at the IAU MDC with the mean values of the same parameters of each found
group. We compared as well the mean orbits, radiant positions and geocentric
velocities with newly meteor showers found in the SonotaCo (2007-2009) and CAMS
(2010-2011) databases (\cite{Rudawska2014}). The mean values of the orbital elements
and other parameters of each group were obtained as a weighted arithmetic mean,
where the weight was determined by $(1-D_i^2/D_c^2)$ (\cite{Welch2001}).

The results obtained from this procedure are given in Table~\ref{tab1} and
Table~\ref{tab2}. Table \ref{tab1} contains 22 showers for which the identification
was certain, i.e. the parameters of which agree well with those from the IAU MDC
list. Another 18 showers, for which the comparison showed quite considerable
differences in some parameters, and thus making their identification questionable,
are shown in Table~\ref{tab2}. For instance, the difference in solar longitude or
right ascension of radiant position reaches 10$^\circ$, while the difference in
eccentricities and perihelion distances, probably due to their high geocentric
velocity, is greater than 0.1. In the EDMOND database (as of August 2013) we could
not identify 25 \textit{pro tempore} meteor showers. The reason is either the number
of orbits in the particular showers was insufficient (less than 5) or the
differences between some of the compared parameters were too big (larger than in
Table~\ref{tab2}).

A few low inclined meteor showers (\#449, \#467, \#473, \#475, \#476, \#478) seem to
be represented as separate branches, where one (or both) of the branch includes from
1 to 3 members. However, as the amount of meteors in the Northern and/or Southern
branch is small ($<$5), and there is no evident splitting, in those cases we
considered such shower as one meteor shower. Therefore, we added (or subtracted) 180
degrees to the angular elements ($\omega$, $\Omega$) of the smaller branch, and then
the weighted mean of $\omega$ and $\Omega$ of the shower was calculated.

\begin{table}
\caption{Mean values of the parameters: solar longitude ($L_S$), radiant position
$(RA,Dc)_{2000}$, geocentric velocity ($V_g$), orbital elements and ($D$) --
Southworth-Hawkins criterion of reliably identified \textit{pro tempore} showers from the
IAU MDC in the database EDMOND. $N$ -- number of meteors. In the second line of each
shower, there are Standard Deviations.} \centering \scriptsize
\begin{tabular}{|r|r|r|r|r|r|r|r|r|r|r|r|}
\hline \multicolumn{1}{|c}{Shower} & %
\multicolumn{1}{|c}{$L_S$} & %
\multicolumn{1}{|c}{$RA$}  & %
\multicolumn{1}{|c}{$Dec$} & %
\multicolumn{1}{|c}{$V_g$}  & %
\multicolumn{1}{|c}{$q$ [AU]} & %
\multicolumn{1}{|c}{$e$}      & %
\multicolumn{1}{|c}{$\omega$ [$^\circ$]} & %
\multicolumn{1}{|c|}{$\Omega$ [$^\circ$]} & %
\multicolumn{1}{|c|}{$i$ [$^\circ$]} & %
\multicolumn{1}{|c|}{$N$} & %
\multicolumn{1}{|c|}{$D_{SH}$}            \\
\hline
448 AAL &  14.4 & 219.7 & -13.0 & 37.70 & 0.097 & 0.945 & 329.6 &  14.4 &   6.7 &   8 & 0.05 \\
$\pm$   &   6.0 &   3.2 &   1.6 &  2.35 & 0.028 & 0.025 &   5.4 &   6.0 &   3.5 &     & 0.04 \\
&&&&&&&&&&&\\[-4pt]
449 ABS &   7.3 & 166.5 &   5.5 & 14.65 & 0.844 & 0.658 &  52.4 & 187.3 &   0.7 &   5 & 0.04 \\
        &   5.1 &   2.4 &   2.7 &  0.82 & 0.023 & 0.050 &   4.6 &   5.1 &   0.8 &     & 0.04 \\
&&&&&&&&&&&\\[-4pt]
456 MPS &  61.7 & 243.7 & -10.5 & 24.63 & 0.541 & 0.790 & 273.2 &  61.7 &   9.0 &  26 & 0.05 \\
        &   4.4 &   2.9 &   1.4 &  1.12 & 0.040 & 0.025 &   4.8 &   4.4 &   1.2 &     & 0.03 \\
&&&&&&&&&&&\\[-4pt]
458 JEC &  83.0 & 315.5 &  33.1 & 52.11 & 0.911 & 0.888 & 218.7 &  83.0 &  95.4 &  10 & 0.05 \\
        &   1.9 &   1.4 &   1.0 &  0.80 & 0.008 & 0.047 &   1.4 &   1.9 &   1.1 &     & 0.03 \\
&&&&&&&&&&&\\[-4pt]
460 LOP &  85.9 & 257.6 &  -5.4 & 19.62 & 0.722 & 0.724 & 251.8 &  85.9 &  10.3 &  27 & 0.05 \\
        &   3.7 &   2.2 &   2.4 &  1.13 & 0.037 & 0.032 &   5.0 &   3.7 &   1.0 &     & 0.03 \\
&&&&&&&&&&&\\[-4pt]
462 JGP & 120.5 & 263.5 &  13.3 & 62.31 & 0.484 & 0.922 & 275.2 & 120.5 & 149.4 &  12 & 0.06 \\
        &   3.5 & 161.1 &   1.1 &  0.57 & 0.040 & 0.034 &   4.2 &   3.5 &   1.7 &     & 0.03 \\
&&&&&&&&&&&\\[-4pt]
463 JRH & 125.8 & 265.9 &  36.2 & 14.18 & 0.982 & 0.553 & 204.5 & 125.8 &  19.7 &   8 & 0.05 \\
        &   5.2 &   2.8 &   2.3 &  0.97 & 0.012 & 0.046 &   5.4 &   5.2 &   1.5 &     & 0.03 \\
&&&&&&&&&&&\\[-4pt]
465 AXC & 136.1 &   4.7 &  48.9 & 54.72 & 0.898 & 0.843 & 221.4 & 136.1 & 104.2 &  14 & 0.06 \\
        &   2.1 &   3.0 &   1.3 &  0.63 & 0.015 & 0.049 &   2.7 &   2.1 &   1.2 &     & 0.03 \\
&&&&&&&&&&&\\[-4pt]
466 AOC & 136.8 &  29.0 &   0.9 & 65.84 & 0.696 & 0.901 &  70.2 & 316.8 & 159.8 &   6 & 0.05 \\
        &   4.5 &   3.5 &   2.1 &  0.56 & 0.025 & 0.048 &   2.9 &   4.5 &   3.6 &     & 0.05 \\
&&&&&&&&&&&\\[-4pt]
467 ANA & 139.5 & 318.1 & -12.2 & 21.35 & 0.612 & 0.752 & 265.6 & 139.5 &   2.6 &  23 & 0.06 \\
        &   3.2 &   2.1 &   2.1 &  1.39 & 0.037 & 0.037 &   4.2 &   3.2 &   1.6 &     & 0.03 \\
&&&&&&&&&&&\\[-4pt]
474 ABA & 147.9 & 301.3 &   4.5 & 15.07 & 0.860 & 0.676 & 230.9 & 147.9 &  10.1 &  11 & 0.05 \\
        &   6.6 &   1.9 &   2.4 &  1.49 & 0.041 & 0.053 &   7.4 &   6.6 &   1.0 &     & 0.03 \\
&&&&&&&&&&&\\[-4pt]
477 SRP & 177.1 & 345.9 &   5.1 & 18.41 & 0.699 & 0.699 & 254.8 & 177.1 &   5.8 &  15 & 0.06 \\
        &   4.2 &   1.8 &   2.0 &  1.51 & 0.046 & 0.043 &   6.6 &   4.1 &   1.1 &     & 0.03 \\
&&&&&&&&&&&\\[-4pt]
478 STC & 170.7 & 315.3 & -13.3 & 10.19 & 0.927 & 0.561 & 218.3 & 170.7 &   1.1 &   6 & 0.05 \\
        &   8.1 &   1.9 &   3.5 &  1.17 & 0.024 & 0.041 &   6.6 &   8.1 &   0.9 &     & 0.03 \\
&&&&&&&&&&&\\[-4pt]
479 SOO & 185.7 &  80.4 &  10.6 & 66.87 & 0.792 & 0.876 &  56.5 &   5.7 & 156.5 &  20 & 0.07 \\
        &   3.1 &   2.2 &   1.6 &  0.69 & 0.031 & 0.049 &   4.6 &   3.1 &   2.9 &     & 0.03 \\
&&&&&&&&&&&\\[-4pt]
480 TCA & 204.1 & 135.1 &  29.2 & 67.31 & 0.808 & 0.839 & 125.9 & 204.1 & 158.0 &  18 & 0.06 \\
        &   3.5 &   3.1 &   1.2 &  0.52 & 0.023 & 0.044 &   3.8 &   3.5 &   2.2 &     & 0.03 \\
&&&&&&&&&&&\\[-4pt]
497 DAB & 261.8 & 210.6 &  22.9 & 59.47 & 0.690 & 0.967 & 113.1 & 261.8 & 113.6 &   5 & 0.03 \\
        &   0.7 &   1.1 &   1.2 &  0.31 & 0.025 & 0.021 &   3.3 &   0.7 &   0.5 &     & 0.05 \\
&&&&&&&&&&&\\[-4pt]
500 JPV & 288.2 & 221.9 &   1.2 & 65.05 & 0.657 & 0.866 & 106.6 & 288.2 & 146.5 &   8 & 0.05 \\
        &   3.4 &   2.4 &   1.4 &  0.92 & 0.028 & 0.056 &   3.1 &   3.4 &   2.6 &     & 0.04 \\
&&&&&&&&&&&\\[-4pt]
502 DRV & 253.2 & 185.1 &  12.3 & 68.18 & 0.776 & 0.920 & 123.8 & 253.2 & 154.8 &   7 & 0.05 \\
        &   4.0 &   3.2 &   1.7 &  0.82 & 0.024 & 0.051 &   4.1 &   4.0 &   2.6 &     & 0.04 \\
&&&&&&&&&&&\\[-4pt]
508 TPI & 146.5 & 351.5 &   4.0 & 38.01 & 0.102 & 0.951 & 328.0 & 146.5 &  21.1 & 143 & 0.06 \\
        &   4.7 &   3.3 &   2.0 &  1.49 & 0.021 & 0.014 &   3.9 &   4.7 &   3.0 &     & 0.03 \\
&&&&&&&&&&&\\[-4pt]
529 EHY & 258.2 & 134.1 &   2.4 & 61.72 & 0.362 & 0.951 & 107.4 &  78.2 & 143.0 &  18 & 0.07 \\
        &   3.1 &   2.6 &   1.0 &  0.97 & 0.029 & 0.031 &   4.0 &   3.1 &   1.8 &     & 0.03 \\
&&&&&&&&&&&\\[-4pt]
530 ECV & 304.9 & 193.9 & -18.6 & 67.39 & 0.790 & 0.813 &  56.0 & 124.9 & 157.9 &   6 & 0.05 \\
        &   3.7 &   3.2 &   1.7 &  0.49 & 0.038 & 0.026 &   5.8 &   3.7 &   3.9 &     & 0.04 \\
&&&&&&&&&&&\\[-4pt]
546 FTC & 144.1 &  30.2 &  67.4 & 52.20 & 1.009 & 0.868 & 173.0 & 144.1 &  95.4 &  14 & 0.06 \\
        &   2.8 &   4.3 &   1.5 &  1.10 & 0.002 & 0.058 &   2.3 &   2.8 &   2.2 &     & 0.04 \\
\hline
\end{tabular}
\normalsize \label{tab1}
\end{table}

\begin{table}
\caption{Mean values of the parameters: solar longitude ($L_S$), radiant position
$(RA,Dc)_{2000}$, geocentric velocity ($V_g$), orbital elements and ($D$) --
Southworth-Hawkins criterion of questionably identified \textit{pro tempore} showers from
the IAU MDC in the database EDMOND. $N$ -- number of meteors. In the second line of
each shower, there are Standard Deviations.} \centering \scriptsize
\begin{tabular}{|r|r|r|r|r|r|r|r|r|r|r|r|}
\hline \multicolumn{1}{|c}{Shower} & %
\multicolumn{1}{|c}{$L_S$} & %
\multicolumn{1}{|c}{$RA$}  & %
\multicolumn{1}{|c}{$Dec$} & %
\multicolumn{1}{|c}{$V_g$}  & %
\multicolumn{1}{|c}{$q$ [AU]} & %
\multicolumn{1}{|c}{$e$}      & %
\multicolumn{1}{|c}{$\omega$ [$^\circ$]} & %
\multicolumn{1}{|c|}{$\Omega$ [$^\circ$]} & %
\multicolumn{1}{|c|}{$i$ [$^\circ$]} & %
\multicolumn{1}{|c|}{$N$} & %
\multicolumn{1}{|c|}{$D_{SH}$}            \\
\hline
451 CAM &  40.6 & 182.7 &  83.2 & 13.02 & 1.000 & 0.517 & 167.9 &  40.6 &  19.0 &   4 & 0.04 \\
$\pm$   &   5.8 &   7.8 &   2.6 &  0.69 & 0.003 & 0.033 &   3.5 &   5.8 &   1.2 &     & 0.04 \\
&&&&&&&&&&&\\[-4pt]
464 KLY & 125.9 & 276.3 &  34.8 & 18.61 & 0.945 & 0.695 & 213.6 & 125.9 &  25.1 &   6 & 0.06 \\
        &   6.8 &   2.2 &   1.9 &  1.32 & 0.018 & 0.043 &   4.3 &   6.8 &   1.5 &     & 0.04 \\
&&&&&&&&&&&\\[-4pt]
468 AAH & 136.3 & 267.8 &  20.6 & 12.47 & 0.977 & 0.631 & 204.4 & 136.3 &  13.5 &   7 & 0.05 \\
        &   7.4 &   2.6 &   2.2 &  1.05 & 0.013 & 0.047 &   5.4 &   7.4 &   1.2 &     & 0.04 \\
&&&&&&&&&&&\\[-4pt]
470 AMD & 144.4 & 254.8 &  58.2 & 18.98 & 1.012 & 0.631 & 178.4 & 144.4 &  29.5 &  17 & 0.07 \\
        &   4.2 &   4.2 &   2.6 &  1.12 & 0.002 & 0.041 &   3.6 &   4.2 &   2.0 &     & 0.03 \\
&&&&&&&&&&&\\[-4pt]
471 ABC & 137.8 & 306.3 & -12.5 & 16.95 & 0.752 & 0.676 & 248.9 & 137.8 &   3.4 &   9 & 0.05 \\
        &   3.8 &   2.2 &   2.4 &  1.42 & 0.035 & 0.043 &   4.6 &   3.8 &   1.5 &     & 0.03 \\
&&&&&&&&&&&\\[-4pt]
472 ATA & 143.8 & 310.3 &  -1.8 & 18.66 & 0.742 & 0.735 & 248.3 & 143.8 &   8.8 &  10 & 0.06 \\
        &   6.3 &   1.9 &   3.6 &  1.29 & 0.046 & 0.044 &   7.1 &   6.3 &   1.5 &     & 0.04 \\
&&&&&&&&&&&\\[-4pt]
473 LAQ & 145.3 & 341.0 &  -5.1 & 31.12 & 0.279 & 0.881 & 303.2 & 145.3 &   4.1 &  20 & 0.07 \\
        &   2.8 &   2.4 &   1.8 &  1.09 & 0.026 & 0.023 &   3.8 &   2.8 &   2.4 &     & 0.04 \\
&&&&&&&&&&&\\[-4pt]
475 SAQ & 157.1 & 330.6 & -10.7 & 21.02 & 0.669 & 0.810 & 255.7 & 157.1 &   0.8 &   8 & 0.06 \\
        &   4.0 &   1.6 &   1.4 &  1.22 & 0.033 & 0.060 &   5.1 &   4.0 &   0.7 &     & 0.05 \\
&&&&&&&&&&&\\[-4pt]
476 ICE & 175.5 &   4.6 &  -0.7 & 26.23 & 0.419 & 0.811 & 107.7 & 355.5 &   2.6 &  21 & 0.07 \\
        &   5.0 &   2.9 &   2.1 &  1.39 & 0.043 & 0.032 &   5.7 &   5.0 &   1.7 &     & 0.03 \\
&&&&&&&&&&&\\[-4pt]
481 OML & 219.7 & 148.5 &  29.1 & 67.13 & 0.892 & 0.793 & 140.7 & 219.7 & 152.1 &   6 & 0.06 \\
        &   3.7 &   3.2 &   1.7 &  1.00 & 0.024 & 0.048 &   5.2 &   3.7 &   2.6 &     & 0.04 \\
&&&&&&&&&&&\\[-4pt]
484 IOA & 233.6 &  28.5 &  15.6 & 14.51 & 0.824 & 0.677 & 233.9 & 233.6 &   1.5 &   5 & 0.05 \\
        &   4.5 &   1.7 &   2.5 &  1.34 & 0.037 & 0.038 &   6.2 &   4.5 &   1.0 &     & 0.05 \\
&&&&&&&&&&&\\[-4pt]
499 DDL & 277.4 & 169.5 &  26.6 & 63.06 & 0.536 & 0.955 & 266.1 & 277.4 & 135.3 &  62 & 0.06 \\
        &   2.8 &   2.5 &   1.4 &  0.85 & 0.022 & 0.044 &   3.0 &   2.8 &   1.6 &     & 0.03 \\
&&&&&&&&&&&\\[-4pt]
531 GAQ &  49.8 & 305.8 &  14.1 & 60.78 & 0.980 & 0.781 & 201.2 &  49.8 & 123.3 &   6 & 0.04 \\
        &   2.7 &   1.9 &   0.5 &  0.30 & 0.012 & 0.048 &   3.7 &   2.7 &   0.9 &     & 0.04 \\
&&&&&&&&&&&\\[-4pt]
533 JXA & 112.6 &  35.0 &   9.2 & 68.85 & 0.863 & 0.939 & 313.8 & 292.6 & 171.8 &  19 & 0.06 \\
        &   6.6 &   4.7 &   2.0 &  0.55 & 0.038 & 0.032 &   6.5 &   6.6 &   2.3 &     & 0.03 \\
&&&&&&&&&&&\\[-4pt]
537 KAU & 207.4 &  90.7 &  32.0 & 64.98 & 0.483 & 0.965 & 272.9 & 207.4 & 160.4 &  10 & 0.06 \\
        &   4.6 &   4.4 &   1.1 &  1.01 & 0.045 & 0.040 &   5.6 &   4.6 &   2.5 &     & 0.04 \\
&&&&&&&&&&&\\[-4pt]
538 FFA & 215.1 &  50.9 &  30.2 & 38.12 & 0.187 & 0.957 & 311.3 & 215.1 &  24.1 &   5 & 0.05 \\
        &   5.8 &   4.8 &   2.7 &  1.87 & 0.021 & 0.029 &   4.0 &   5.8 &   3.3 &     & 0.05 \\
&&&&&&&&&&&\\[-4pt]
545 KCA & 156.4 &   8.6 &  49.4 & 51.36 & 0.685 & 0.925 & 250.7 & 156.4 &  93.6 &   4 & 0.05 \\
        &   1.9 &   2.5 &   1.4 &  0.94 & 0.019 & 0.049 &   2.0 &   1.9 &   1.3 &     & 0.04 \\
&&&&&&&&&&&\\[-4pt]
547 KAP & 137.6 &  43.9 &  45.6 & 63.53 & 0.976 & 0.852 & 157.0 & 137.6 & 132.0 &  11 & 0.06 \\
        &   2.1 &   2.6 &   1.4 &  0.50 & 0.009 & 0.040 &   3.0 &   2.1 &   2.2 &     & 0.03 \\
\hline
\end{tabular}
\normalsize
\label{tab2}
\end{table}

\section{Conclusions}

In the work, the European viDeo MeteOr Network Database (EDMOND) is introduced. Its
4th version contains 83\,369 video meteor orbits. The Database was created thanks to
the broad international cooperation of several European national networks and the
International Meteor Organization Video Meteor Network, with the aim of connecting
observers within a wide area. This has made it possible to combine those
observations which otherwise would have stayed as single-station data.

We expect to use this expanding database, particularly to study minor streams. The
first results are here presented. Using the \textit{radiant-$V_g$} method we
identified 267 meteor showers of the IAU MDC, where 67 of them are established
showers and 200 are showers from the working list.

Making a more detailed analysis of \textit{pro tempore} showers from the IAU MDC
working list, we determined their orbital elements, radiant positions, geocentric
velocities and solar longitudes. The results were divided into two groups, based on
a comparison of the mean values with those available in the IAU MDC. Table
\ref{tab1} contains 22 showers of the 65 \textit{pro tempore} showers in the working
list of the IAU MDC (August 2013), identification of which was clear and reliable.
Identification of 18 meteor showers listed in Table \ref{tab2} is questionable, as
some their parameters differ considerably from those at the IAU MDC.

This work showed that the EDMOND database is able to provide relevant data
convenient for the confirmation of meteor showers from the working list of the IAU
MDC, which can improve their orbital and geophysical parameters.

\begin{acknowledgments}
This work was supported by the Slovak Research and Development Agency, project No.
APVV-0517-12.
\end{acknowledgments}


\begin{thebibliography}{}
%

\bibitem[{Andrei\'{c} and \v{S}egon (2010)}]{Andreic2010}
Andrei\'{c}, \v{Z}., \v{S}egon, D., 2010, In Proceedings of the International Meteor
Conference, \v{S}achti\v{c}ka, Slovakia, 18--21 September 2008, 16
%
\bibitem[Brown et al. (2010)]{Brown2010}
Brown, P., Weryk, R.~J., Kohut, S., Edwards, W.~N., Krzeminski, Z.,
2010, WGN, Journal of the International Meteor Organization, 38, 25
%
\bibitem[Cooke and Moser (2012)]{Cooke2012}
Cooke, W.~J., Moser, D.~E., 2012, In Proceedings of the International Meteor
Conference, Sibiu, Romania, 15--18 September 2011, 9
%
\bibitem[Hajdukov\'{a} (2013)]{Hajdukova2013}
Hajdukov\'{a}, M.~Jr., 2013, In Proceedings of the International Meteor Conference,
Pozna\'{n}, Poland, 22--25 August 2013, in press
%
\bibitem[Hajdukov\'{a} et al. (2014a)]{Hajdukovaetal2014a}
Hajdukov\'{a}, M.~Jr., Korno\v{s}, L., T\'{o}th, J. 2014a, Meteoritics Planet. Sci.,
49, 63
%
\bibitem[Hajdukov\'{a} et al. (2014b)]{Hajdukova2014b}
Hajdukov\'{a}, M.~Jr., Korno\v{s}, L., T\'{o}th, J. 2014b, in this Proceedings, p ??
%
\bibitem[Jenniskens et al. (2011)]{Jenniskens2011}
Jenniskens, P., Gural, P. S., Dynneson, L., Grigsby, B. J., Newman, K. E., Borden, M.,
Koop, M., Holman, D., 2011, Icarus, 216, 40
%
\bibitem[Jopek and Ka\v{n}uchov\'{a} (2014)]{Jopek2014}
Jopek, T.~J., Ka\v{n}uchov\'{a}, Z., 2014, in this Proceedings, p ??
%
\bibitem[Korno\v{s} et al. (2013a)]{Kornos2013a}
Korno\v{s}, L., Koukal, J., Piffl, R., T\'{o}th, J., 2013a, In Proceedings of the
International Meteor Conference, Canary Islands, Spain, 20--23 September 2012, 21
%
\bibitem[Korno\v{s} et al. (2013b)]{Kornos2013b}
Korno\v{s}, L., Koukal, J., Piffl, R., T\'{o}th, J., 2013b, In Proceedings of the
International Meteor Conference, Pozna\'{n}, Poland, 22--25 August 2013, in press
%
\bibitem[Lindblad et al. (2005)]{Lindblad2005}
Lindblad, B.~A., Neslu\v{s}an, L., Porub\v{c}an, V., Svore\v{n} J.,2005 Earth, Moon,
Planets 93, 249
%
\bibitem[Molau (1999)]{Molau1999}
Molau, S., 1999, In Proceedings of the International Conference Meteoroids 1998,
Tatranska Lomnica, Slovakia, 17--21 August 1998, 131
%
\bibitem[Molau (2014)]{Molau2014}
Molau, S., 2014, in this Proceedings, p ??
%
\bibitem[Pujols et al. (2013)]{Pujols2013}
Pujols, P., Ambl{\`a}s, M., Trigo-Rodr{\'{\i}}guez, J.~M., Dergham, J., Madiedo,
J.~M., Montany{\`a}, J., van der Velde, O., 2013, Lunar and Planetary Institute
Science Conference Abstracts, 44, 2054
%
\bibitem[Rudawska and Jenniskens (2014)]{Rudawska2014}
Rudawska, R., Jenniskens, P., 2014, in this Proceedings, p ?
%
\bibitem[SonotaCo (2009)]{SonotaCo2009}
SonotaCo, 2009, WGN, Journal of the International Meteor Organization, 37, 55
%
\bibitem[SonotaCo et al. (2010)]{SonotaCo2010}
SonotaCo, Molau, S., Koschny, D., 2010, European Planetary Science Congress 2010,
798
%
\bibitem[{Southworth and Hawkins (1963)}]{SouthworthHawkins1963}
Southworth R.~B., Hawkins G.~S., 1963, Smithsonian Contrib. Astrophys. 7, 261
%
\bibitem[{T\'{o}th et al. (2011)}]{Toth2011}
T\'{o}th, J., Korno\v{s}, L., Vere\v{s}, P., \v{S}ilha, J., Kalman\v{c}ok, D., Zigo,
P., Vil\'{a}gi, J., 2011, Publ. Astron. Soc. Japan 63, 331
%
\bibitem[{Vere\v{s} and T\'{o}th (2010)}]{Veres2010}
Vere\v{s}, P. and T\'{o}th, J., 2010, WGN, Journal of the International Meteor
Organization, 38, 54
%
\bibitem[Welch (2001)]{Welch2001}
Welch, P.~G., 2001, Mon. Not. R. Astron. Soc. 328, 101

\normalsize
\end{thebibliography}
\end{document}